# TOWARDS INCREASING TRUST IN EXPERT EVIDENCE DERIVED FROM MALWARE FORENSIC TOOLS

Ian Kennedy, Arosha Bandara, Blaine Price
The Open University
School of Computing and Communications
Milton Keynes MK7 6AA, U.K
ian.m.kennedy@open.ac.uk

**ABSTRACT**

Following a series of high profile miscarriages of justice in the UK linked to questionable expert evidence, the post of the Forensic Science Regulator was created in 2008. The main objective of this role is to improve the standard of practitioner competences and forensic procedures. One of the key strategies deployed to achieve this is the push to incorporate a greater level of scientific conduct in the various fields of forensic practice. Currently there is no statutory requirement for practitioners to become accredited to continue working with the Criminal Justice System of England and Wales. However, the Forensic Science Regulator is lobbying the UK Government to make this mandatory. This paper focuses upon the challenge of incorporating a scientific methodology to digital forensic investigations where malicious software ('malware') has been identified. One aspect of such a methodology is the approach followed to both select and evaluate the tools used to perform dynamic malware analysis during an investigation. Based on the literature, legal, regulatory and practical needs we derive a set of requirements to address this challenge. We present a framework, called the 'Malware Analysis Tool Evaluation Framework' (MATEF), to address this lack of methodology to evaluate software tools used to perform dynamic malware analysis during investigations involving malware and discuss how it meets the derived requirements.

## 1. INTRODUCTION

It is not uncommon during cybercrime investigations to discover malware. In R v Oliver (R v Oliver [2016] EWCA Crim 1053) a 'Trojan defence' was offered to account for the presence of indecent images on a computer, while in R v Dan (R v Dan [2019] EWCA Crim 1985) it was argued that a "malware bug…potentially affected its operation."

In the UK, the Criminal Procedure Rules (Ministry of Justice, 2015) stipulate that digital forensic practitioners have a duty to assist the court in their understanding of the evidence tendered. Therefore, they also have a duty to identify the capabilities of any malware identified during their investigation and make a determination as to the bearing it has (if any) on the conclusions reached.

To undertake this duty (and hence form an opinion on the impact of any malware), the forensic practitioner is reliant on their tools, skills and knowledge of malware to detect, identify and study the behaviour of any identified malware. However, practitioners are relying on anecdotal or otherwise limited scientific principles to form their conclusions on such impact (Kennedy, 2017). This is contrary to scientifically based decision making, which is "an assumed trait of the practitioner, rather than a formally taught competency" (Horsman, 2019a).

This is in part due to the lack of an established methodology for malware forensics undertaken as part of a criminal investigation. The unpredictable nature of malware means this lack of an established methodology could violate legislation such as the Computer Misuse Act (1990). It could also violate technically led best practice guidelines (Williams, 2012) and more recently the quality focused Codes of Practice and Conduct (Forensic Science

Regulator, 2020b), hereafter referred to as "the Codes".

One aspect of such a methodology is the approach followed to both select and evaluate the tools used to analyse the malware and the artefacts it produces. Broadly speaking, there are two approaches used to study malware: dynamic and static. The former monitors malware that is operational. The latter examines malware in a passive state, by studying the underlying code, for example. Tools will generally support one (sometimes both) of these approaches. Existing methods to evaluate tools in a conventional digital forensic examination include 'dual-tool verification' promoted by the Association of Chief Police Officers (ACPO) (Clarke, 2009). Arguments that a tool has been widely accepted in case law (Guidance Software Inc., 2014) are open to challenge when examined at a statistically significant scale (Kennedy, 2017) and limited in their utility (Horsman, 2019b).

This work provides a foundation to determine if a systematic basis for trusted practice could be established for evaluating malware artefact detection tools used within a forensic investigation. The contributions of the work are to (a) identify the legislative, technical and quality requirements of malware forensic practice; (b) provide a framework to address these requirements.

The structure of this paper is as follows: Section 2 explores the background and related work, while section 3 derives the requirements for undertaking malware forensics. Section 4 describes a framework to address these requirements. Section 5 reflects on the framework and the extent to which the requirements have been addressed, while section 6 draws conclusions and identifies further work.

## 2. BACKGROUND AND RELATED WORK

The digital fields of *malware* and *forensics* are increasingly being combined to describe what Malin et al. (2008) term *malware forensics*. As a field of study in its own right, universities are now beginning to offer malware forensics either as a whole course/module (University of Portsmouth, 2019) or as part of related modules, such as Digital Forensics (University of London, 2020).

The original motivation for this work arose from the realisation that digital forensic practitioners were conducting malware forensic investigations in a largely anecdotal manner. This may be in part due to the fact that there is little published material establishing a scientific basis for procedures applied to conducting a malware forensic investigation; and more specifically for evaluating the tools to do so. Liu et al. (2017) applied malware ontology techniques to assist investigators by providing a means to categorise malware behaviour in terms of one of five broad categories. The definitions applied lack a rationale and do not address malware that occupies more than one category. Furthermore, while helpful to a lay audience, the approach does not assist investigators in understanding the impact (if any) of any malware found on a computer under investigation.

The use of malware forensics is cited by Kim et al. (2014) who presented a model to investigate fraud using "malware forensic" techniques. Provataki and Katos (2013) offered a framework that extends the functionality of the Cuckoo sandbox (Cuckoo Foundation, 2016) to understand malware's behaviour but not to evaluate the tools used to study such behaviour. Shosha et al. (2013) explored the limitations of dynamic malware analysis techniques for digital investigations, which are highlighted and propose a methodology to analyse malicious code running in forensically acquired computer memory. However, the methodology proposed is only applicable to analysis of code running in memory and is not based on any formal requirements analysis.

A malware analysis approach was proposed by Ianelli et al. (2007) who suggested that the presence of malware can be addressed by examination of the network traffic logs. However, this suggestion assumes that such logs are more likely to be found in a corporate than domestic environment. Hence, a suspect accused of committing an offence via their home router will typically have far fewer logs and/or detail to assist their defence than in a commercial environment, where there would likely be more sophisticated logging available.

Malin et al. (2008) presented one of the few books on malware forensics, more recently split into separate Windows (2012) and Linux (2013) editions. Carvey (2012) also provided some coverage of the topic across two chapters from an investigative perspective, as part of a more general digital forensics discussion. Each of these texts presents a collection of tools and techniques to address various aspect of analysis, but none attempt to develop and evaluate a general-purpose framework for malware analysis or a rigorous, scientific means to evaluate the tools used.

The lack of a formalised approach means it is also not uncommon to find tools not specifically designed for forensic use being deployed. Hughes and Varol (2020) argued that the use of malware scanners, employed to identify malware in a forensic investigation, will not meet all possible functional requirements. For example, such tools are not designed to detect malware that previously existed on a machine and is now located in areas such as slack space, unused partitions and deleted files. Thus, the validity of tools (and hence any resulting conclusions) can be undermined by their application to scenarios for which they were not designed.

Perhaps more significantly, the lack of a formalised approach means that court proceedings involving malware may not be properly investigated. Such cases will inevitably become a candidate for miscarriages of justice, as the court would be forming a judgment without being fully informed of the facts. Some of the challenges that can impair an investigation involving malware are explored below.

## 2.1 The Trojan Defence

Separating user actions from those of malicious software is the fundamental objective when investigating the Trojan defence, where a defendant can claim the illegal activity recorded on a device is the result of malware and not their own actions. Brown (2015) highlighted the Trojan defence as one of several tactics used by counsel to raise doubt as to the authenticity of the electronic evidence presented to court. Bowles and Hernandez-Castro (2015) highlighted "clear and obvious mistakes" with regard to Trojan defence cases in a study covering a 10 year period.

The problem of attribution is anticipated to become more challenging in the near future with the nefarious use of artificial intelligence (AI) to enhance malware. Thanh and Zelinka (2019) warn of an upcoming 'AI-powered malware era', citing proof of concept work that demonstrates that 'computational intelligence could be used to enhance malware.' This warning is echoed by Truong et al. (2020) who identify deep learning techniques being applied to malware.

Bikeev et al. (2019) explored the challenges of applying *mens rea* to malicious AI and Bahnsen et al. (2018) developed an algorithm to enhance AI to be more effective during malicious phishing attacks. Alongside malware and AI, doubt in the reliability of digital evidence can also originate from the methodologies followed by forensic practitioners. Perhaps the most significant of these is an over dependence on anecdotal experience when reaching conclusions.

## 2.2 Repeated Confirmation

Sceptical digital forensic practitioners may defer to their anecdotal experience to argue that they are "yet to see an example" (McLinden, 2009) and similarly that they "haven't seen a single case" (Douglas, 2007) of malware attributed to the downloading of indecent images of children. Similarly, the results from mainstream digital forensic tools have been accepted "based solely on the reputation of the vendor" (Garfinkel, Farrell, Roussev, & Dinolt, 2009). Such arguments are formulated on inductive reasoning, derived from repeated confirmation. Although useful to develop hypothesises, inductive reasoning cannot be used to test scientific theory (Levitin, 2016). There are also challenges in the processing and reasoning that is applied to expert evidence.

## 2.3 Ubiquitous Problems With Expert Evidence

Challenges with the evidence include experts who step outside of their own expertise. The now infamous trials of R v Clark [2003] EWCA Crim 1020, R v Cannings [2004] EWCA Crim 1 and R v Patel [2003] provide examples of where the defence expert, Professor Sir Roy Meadows, made a number of claims that had "no statistical basis" (Royal Statistical Society, 2001). Following these events, the Law Commission's review (2011) of expert evidence in criminal trials called for a move to incorporate a greater level of scientific principles and provenance in expert evidence.

Challenges also arise with failures to find and/or disclose evidence correctly. Bowcott (2018) cites problems at a series of criminal trials where digital evidence was either not found nor passed to the defence team during disclosure.

Problems with expert evidence are not limited to the UK alone. Edmond and Vulle (2014) examined the use of forensic science evidence in trials and concluded three separate criminal justice systems (United States, Switzerland and Australia) each failed to identify "deep structural and endemic problems with many types of forensic science".

Edmond and Vulle (2014) go on to argue that these problems extend to the use of language by experts, stating that the "expressions used by analysts are not empirically based". This is echoed by Adam (2016) who argues that language used for conclusions such as 'it is likely' is based on posterior probabilities and so implies probabilistic support to the conclusion. However, such phrases almost never provide any detail on how the likelihood has been reached. Such a conclusion could be based on unreported properties of items considered or entirely subjective. Similarly, Adam (2016) goes on to challenge the phrase 'is consistent with', which states some (unknown) degree of similarity between two things. Typically, either no alternative sources are given or a sense of how common the 'consistent' features are in the wider population. The misuse of language in this way may be linked to a lack of understanding of the underlying scientific principles by practitioners.

## 2.4 Lack Of Scientific Principles

Casey (2019) argues that digital forensics is distinct from forensic science "despite over a decade of effort to break down the borders between them." He goes on to argue that some practitioners accept results as "factual", failing to recognise the need for scientific treatment. This leads to problems in recognising and reporting error rates, quantifying levels of confidence in findings or reporting on alternative interpretations of findings. Christensen et al. (2014) argues practitioners appear to have either misunderstood the term 'error' or lacked the skills to correctly apply statistics or the scientific method. They add, practitioners have reportedly claimed either that there is a zero error rate, that such an error rate cannot be estimated or that practitioners have attempted to "calculate error rates *post facto*".

The challenges faced by the digital forensics field are exacerbated within the relatively young malware forensics field by issues such as malware routinely obfuscating its true intentions and hindering attempts to analyse it (Wagener, Dulaunoy, & Engel, 2008). There is therefore a level of uncertainty associated with any conclusions drawn from malware forensics. This uncertainty can be used to raise reasonable doubt about the true nature and intentions of malware. Furthermore, the complexity of the subject matter and the specialist skills required to study it (e.g.: reverse engineering & assembly language) may make the specialty less accessible to practitioners.

This perceived lack of scientific principles arguably also informs the methods used by practitioners to test software tools when attempting to evaluate the reproducibility of the results reported.

## 2.5 Reproducibility flaws

Techniques such as *dual-tool verification* are used by practitioners to "confirm result integrity during analysis" (Forensic control, 2011). To state that a two observations "confirm" a finding is a bold claim and little more than an example of repeated confirmation. It also fails to consider the possibility that both tools are incorrect and simply (erroneously) in agreement (Beckett & Slay, 2007). Hence use of Dual-tool verification in this way cannot confirm a result, but it can corroborate it on a statistically insignificant scale, identifying any discrepancies. An example of this arose is in the trial of Anthony Casey (*State of Florida v. Casey Marie Anthony*, 2011) where a discrepancy was identified between two Internet history tools used to produce expert testimony.

## 2.6 Emerging statutory requirements

The importance of establishing quality standards for forensic science practice has become increasingly apparent in recent years. Interest in this has been expressed throughout the European Union (EU) by forensic institutions, the scientific community, as well as judicial and political stakeholders (van Ruth & Smithuis, 2019).

Consequently, a European Council Framework Decision was passed requiring all member states to set up systems to accredit their forensic service providers carrying out laboratory activities. The UK's response to this was to form the post of the Forensic Science Regulator (FSR). The FSR's Codes (2020b) place an obligation on practitioners to gain accreditation that is mapped to the international standard ISO/IEC 17025 and "embed a systematic approach to quality" (Tully et al., 2020).

### 2.7 Summary

Given the reasons for the appointment of a FSR and the push to make the quality standards statutory, the issues identified currently undermine the trust that can be placed in findings tendered in criminal proceedings.

The production of digital evidence therefore requires the use of reliable tools and competent practitioners who use appropriate scientific language to instil the conditions for trusted practice, particularly when using tools to analyse malware as part of a digital forensic investigation.

Given malware forensics is an emerging field, there is a need develop a scientific methodology to formalise the practice and hence underpin trusted practice in the field. In particular, a methodology to quantifiably evaluate tools used as part of a malware forensic investigation needs to be established. The next section will focus on identifying the requirements for such a methodology.

## 3. REQUIREMENTS FOR CONDUCTING MALWARE FORENSICS

The elicitation of requirements can be done from multiple sources and not simply stakeholders alone (Burnay, 2016). In his study, Burnay (2016) found that eliciting requirements from existing documentation to be significantly faster than the use of stakeholders. It was also found that there were several examples of where stakeholders made statements that conflicted with formally documented requirements and so were either incorrect or simply misunderstood by stakeholders.

The use of authoritative document sources such as legislation and regulatory guidance already embody the requirements of stakeholders and so we took a document study approach to derive the requirements for conducting malware forensics.

Having considered the methodology, the issue of addressing trusted practice in malware forensics could begin by better exploring what is meant by trust. This can be defined as "willingly acting without the full knowledge needed to act" (Duranti & Rogers, 2012). In the context of the Criminal Justice System involving expert evidence, this translates to a Court coming to a decision on the reliability of such evidence based upon two forms of trust: the expert and their evidence.

The former concerns the expert's knowledge and skills, as well as their ability to communicate these effectively and fairly. A shortfall in one of these areas can impact on the interpretation of the evidence or its probative value (see *Problems with expert evidence,* above).

The latter relates to the trust placed on the reliability of the evidence itself. Since the repeal of section 69 of the Police and Criminal Evidence Act 1984 the Crown Prosecution Service (CPS) issued guidance stating that any evidence produced by a computer is presumed to be reliable (CPS, 2014). However, the formation of the FSR and the associated Codes (2020b) indicate that expert evidence has transitioned from an assumed, innate trust to one that is now externally validated. The CPS, FSR and practitioners themselves are all stakeholders in this process, each having their own requirements.

### 3.1 Legal Requirements

As with digital forensic practice in general, the legal requirements for malware forensic practice can be divided into lawful practice and admissibility. For the former, the primary risks are in the handling of the malware files themselves and potential breaches of the computer misuse and/or data protection legislation. The latter requires that any output of a tool used to analyse malware which is tendered as evidence must be admissible. This broadly translates to a person familiar with the expected output of a computer being available to give evidence (Lloyd, 2020). However, few people would be familiar with the expected output of a tool used to analyse malware, which typically produces unpredictable artefacts.

Guidance on expert evidence from the CPS (2019) states that expert evidence will be admissible under common law where:

- It will be of assistance to the court
- The expert has relevant expertise

- The expert is impartial
- The expert evidence is reliable

The first of these requirements concerns the forming of a judgement on the probative value of the evidence tendered, whilst the second and third concern a judgement on the expert. The last requirement concerns both the evidence and the manner in which it was produced. In their guidance, the CPS define reliable evidence in terms of it having a "scientific basis". This indicates a scientific methodology, characterised by attributes such as repeatability, reproducibility, a testable hypothesis, controllability and being unbiased.

Further to the above, the CPS acknowledge that novel techniques are frequently used in a fast-evolving technology discipline and defer to the recommendations of R v Lundy ([2013] UKPC 28), see Table 1, referred to hereafter as the "Lundy Guidelines":

Table 1 : R v Lundy Guidelines

| # | Guideline |
|---|---|
| 1 | Whether the theory or technique can be or has been tested |
| 2 | Whether the theory or technique has been subject to peer review and publication |
| 3 | The known or potential rate of error or the existence of standards |
| 4 | Whether the theory or technique used has been generally accepted |

In response to the some of the problems outlined in the introduction, regulatory codes of practice have been introduced.

### 3.2 Regulatory requirements

Practitioners tendering expert evidence within the criminal justice system are expected to align their practice to regulatory standards, namely the Codes, published by the FSR (2020b). Currently there is no statutory requirement for practitioners to align their work to the Codes, but the FSR is lobbying the UK Government to make this mandatory (Forensic Science Regulator, 2020a).

The Codes stipulate that software tools must be validated (Section 24.1.2 of the Codes) and that an estimate of uncertainty be provided (Section 22 of the Codes). Furthermore, any reference datasets used to test tools against should also be reported (Section 23.4 of the Codes). Each of these requirements is now considered in turn.

*Validation*

The Codes define validation as a means to demonstrate that a "method, process or device is fit for the specific purpose intended". Although not specifically mentioned, the meaning of 'device' could readily be applied to a software device or tool. However, it is not clear how such validation is performed or what metrics should be used to inform a decision how 'fit for purpose' a device is, e.g., accuracy, repeatability, etc.

One measure readily available is that of error, i.e.: the difference between the expected and observed values (Kat & Els, 2012). Malin et al. (2008) point out that the names of artefacts (such as filenames) are typically randomly assigned. In light of this, it is reasonable to expect artefact values to vary much more than the quantity of artefacts produced each time a malware binary is executed. Such behaviour can be validated by repeatedly executing malware and monitoring the quantity of artefacts produced. Furthermore, to quantify and enable statistical analysis of the error, a methodology for testing tools used for malware analysis could examine the difference in artefact quantities, rather than the values themselves.

Validation of a tool measuring artefacts produced by malware is complicated by the fact that malware employs techniques (often termed 'anti-forensic') to obfuscate the truth. Hence 'ground truth' is difficult to establish. One way forward is to compare what is reported by a tool against an independent and trusted source or 'oracle'. This will require the testing methodology to (a) determine the expected value from an independent source and (b) be capable of retrieving the observed number of artefacts from a variety of tools applied to the framework for testing.

*Estimate of uncertainty*

A measure of statistical confidence can contribute to an estimate of uncertainty. One way to calculate this would be to run multiple tests under the same conditions and record the error between the expected and observed numbers of artefacts. The rationale behind this is that the ISO/IEC 17025 Standard (ISO, 2005) upon which the Codes are based, derives its definition of uncertainty from the *Guide to the Expression of Uncertainty in Measurement (GUM)*, produced by the Joint Committee for Guides in Metrology (JCGM) (2008). In this document uncertainty is defined in terms of the "dispersion of the values" associated with an observable quantity. Acknowledging the two components of error (systematic and random), they add that random error "…can usually be reduced by increasing the number of observations". Furthermore, calculating the experimental standard deviation "of the arithmetic mean or average of a series of observations" provides "a measure of the *uncertainty* of the mean due to random effects."

Hence, by running sufficient tests, it would be possible to plot frequency distributions with associated confidence intervals. Similarly, by varying the conditions of tests it would become possible to see the impact of such changes upon the level of uncertainty in the results.

*Reference data sets*

Becket (2010) states there is a "need" for forensic practitioners to demonstrate that "certified reference materials" have been used to evaluate their tools, citing Section 5.6.3.2 of the ISO/IEC 17025 Standard (ISO, 2005). This is not quite accurate as the same section of the Standard states this should be done "where possible". Section 21.2.64(h) of the Codes requires there to be a plan in place for the use of such data. A number of attempts over the years have been made by the scientific community to address the lack of standardised test data; these include test images produced by Carrier (2010), the Computer Forensic Reference Data Sets (CFReDS) project (NIST, 2016), and the Digital Corpora (2017) developed by Garfinkel et al. (2009) as an extensive collection of both fabricated and real data. Although there are also datasets that include malware samples, such as contagion (2020) and VirusShare (2020) these (like those above) are not labelled as being *certified*.

Aside from legal and regulatory requirements, the handling and analysis of malware reveals a number of technical requirements identified by the literature for malware forensics practice.

### 3.3 Practice requirements

The following practice requirements have been identified:

Virtual Machines

Malin et al. (2008) recommend the use of Virtual Machines (VMs), particularly as this provides testing at scale and speed. Hence greater numbers of tests can be performed for repeatability or breadth of testing purposes. To facilitate this, testing should be automated as far as practically possible. In addition, to minimise the risk of malware escaping from a Windows Guest VM (Tank, Aggarwal, & Chaubey, 2019), a Linux host should be used.

In considering the use of virtual machines, it should be noted that malware can detect virtual environments and change its behaviour or even become 'misleading' (Ferrie, 2007). However, it is also noted that in recognition of the ubiquitous use of virtual servers, a shift in malware no longer avoiding virtual environments has also taken place (Wueest, 2014).

Network services

Isolating malware from a network or even the Internet could limit the behaviour exhibited (Deng & Mirkovic, 2018). To counter this, it is a good idea to provide the malware with as many services as possible that it is likely to rely upon, such as SMTP, HTTP and DNS. The use of iNetSim (Hungenberg & Eckert, 2016) is a popular choice in this area. Phu et al. (2019) use it to trap DNS queries from malware under analysis using an iNetSim simulated network; Sikorski & Honig (2012) use it to simulate a broader range of network services; a mixed solution is proposed by Palkmets et al. (2014) who additionally provide a route to the Internet via an onion router network; a malware clustering technique is offered by Fang et al. (2020) who use iNetSim to provide a means to identify the family of malware under analysis.

Vulnerable environment

Alongside network services, vulnerable environments are also key to maximising the behaviour of malware (Szor, 2005). This is echoed by Malin et al. (2008) and Elisan (2015) who goes further and anecdotally promotes the use of "malware friendly" configurations. These include assigning administrator rights to the default user account, disabling auto updates, disabling User Access Control (UAC), setting the Internet browser to the minimum security level, installing commonly exploited software, and creating honeypot files with filenames such as "salaries.xls".

Black box testing

Many of the tools used by digital forensic practitioners, including the mainstream forensic software tools, are closed source (Talib, 2018). Hence there is no access to verify the underlying algorithms used (Horsman, 2019b). Therefore, a black box testing strategy is more viable than a white box testing approach. Furthermore, digital forensic practitioners would have neither the time nor the skills to review source code (Horsman, 2020).

The above requirements have been included in the framework and are summarised in Table 3.

With the requirements identified, the following section will identify the aims of a framework to test tools used for malware forensics in light of the above requirements.

## 4. DESIGN OF THE FRAMEWORK

Having identified the requirements for conducting malware forensics, it follows that the aims for a framework to address these requirements also needs to be determined.

### 4.1 Aims

Malware investigations can use a variety of software tools, some of which make claims to be suited for malware analysis. The proposed framework, referred to as the Malware Analysis Tool Evaluation Framework (MATEF), should provide a mechanism to evaluate these tools by quantifying their ability to detect artefacts produced by real-world malware samples (see Aim 1, Table 3).

Malin et al. (2008) argued that mawlare analysis can be divided into three broad techniques: temporal, relational and functional analysis. Temporal analysis is concerned with the timeline of events surrounding reported activity; while relational analysis refers to the interaction between components of the malware and its environment. Finally, functional analysis relates to the actions the malware is reported to have performed.

The MATEF provides a mechanism to evaluate dynamic analysis software tools. It provides a means to measure the extent to which tools detect the artefacts produced by malware behaviour (see Aim 2, Table 3). On a Windows computer, this behaviour typically manifests itself in the form of file, registry, process and network based artefacts.

Unlike regular software that is largely predictable, malware can be unpredictable in that some behaviour (and hence artefacts) may not be observed. This can happen when the required (and unknown) trigger conditions for a given binary are not met (Nataraj, Karthikeyan, Jacob, & Manjunath, 2011). Thus, the behaviour of malware can be nondeterministic and vary, particularly if it is of a type that communicates with a Command and Control (C&C) server (Akinrolabu, Agrafiotis, & Erola, 2018). Furthermore, malware can include 'measures to impede automatic and manual analyses' (D'Elia, Coppa, Palmaro, & Cavallaro, 2020). Strategies include code obfuscation (Singh & Singh, 2018), detection of debuggers or virtual machines (Chen, Huygens, Desmet, & Joosen, 2016) and deployment of 'split-personality' malware techniques to change the behaviour of code when it is subjected to analysis (Murali, Ravi, & Agarwal, 2020). Such techniques are designed to give misleading results under analysis. Hence mitigation against such risks should be considered when drawing conclusions from the testing of tools used to study malware, see Aim 3, Table 3.

Having identified the aims of the framework, consideration was then given as to how to achieve these aims. Hence, the following section seeks to identify the main components of the framework.

Table 2 : Aims of the framework

| # | Aim |
|---|---|
| 1 | Use real-world malware |
| 2 | Evaluate a tool's ability to detect malware artefacts |
| 3 | Mitigate against anti-forensic techniques |

Table 3 : Proposed requirements

| # | Requirement | Rationale for inclusion |
|---|---|---|
| **Legal Requirements** | | |
| 1 | Handling of malware and what it may access should be controlled. | Compliance with computer misuse and data protection legislation. Controlled access to minimise accidental or deliberate misuse. |
| 2 | Output of tested tool must be admissible. | To comply with admissibility requirements. |
| 3 | Malware analysis tool output must be reliable. | Compliance with requirements for scientific evidence, based upon R v Lundy ([2013] UKPC 28). |
| **Regulatory requirements** | | |
| 4 | Novel methods must be validated. | See section 21.1.5 of the Codes (Forensic Science Regulator, 2020b). |
| 5 | The theory/technique should be peer reviewed or published. | See section 21.2.60 of the Codes (Forensic Science Regulator, 2020b). |
| 6 | The method should be a generally accepted. | See section 28.4.3(e) of the Codes (Forensic Science Regulator, 2020b). |
| 7 | Tools must be validated. | See section 24.1.2 of the Codes (Forensic Science Regulator, 2020b). Implement this by measuring difference of expected and observed numbers of artefacts. This would provide a measure of how close a tool's measurement of a quantity fits with the expected quantity (error). |
| 8 | Estimate of uncertainty. | See section 22 of the Codes (Forensic Science Regulator, 2020b). |
| 9 | Traceability of reference datasets. | See section 21.2.64(h) of the Codes (Forensic Science Regulator, 2020b). |
| **Practice requirements** | | |
| 10 | Use a virtual machine. | Shorter test times, testing at scale, ease of automation and remote control. |
| 11 | Network service provision. | Provide a controlled means to enable the malware to exhibit its behaviour. |
| 12 | Use a vulnerable environment. | Provide a fertile environment to promote 'best case' results for the tools analysed. |
| 13 | Black box testing approach | Recognition of the extensive use of closed source software and the lack of time/skills for practitioners to analyse source code. |
| 14 | Metrics: Determine the expected quantity of artefacts from an independent source. | In the absence of ground truth, provide an independent and authoritative measure to compare a tool against. |
| 15 | Metrics: Read the observed number of artefacts from a variety of tools under test. | Focus upon quantities rather than values to mitigate the manifestation of random values. |

## 4.2 Identifying & selecting the main components of the framework

The MATEF framework includes a number of components to satisfy the aims identified in Table 3. Each of these elements is briefly explored in the following sections, starting with the malware binaries themselves.

*Malware sample source*

To maximise the validity of the evaluation process, real-world malware (a.k.a. malware 'in the wild') is used instead of fabricated malware (see Aim 1, Table 3). The stored malware employs password protected zip files to minimise the contamination risk during handling (see Requirement 1, Table 2). All samples are analysed offline (see Requirement 1, Table 2). Malware can be obtained from any source and imported to the malware library.

*Malware library*

The malware library is a store of malware executables, each accessible through a consistent file naming convention, thus facilitating automation and use of VMs (see Requirement 10, Table 2). Access to this library is restricted to authorised users of the framework only (see Requirement 1, Table 2).

In addition to the malware binary file, information on its expected behaviour should also be stored locally as well (satisfying requirement 14 from Table 2). To be made readily available, this information will be stored in a malware database.

*Malware database*

The malware database stores properties of each malware binary held in the malware library. As a minimum, the details stored include the hash value of the binary and the number of artefacts generated as a result of creating, modifying or deleting files or registry keys. In addition, also stored are the number of ports opened and processes spawned as a result of executing the malware using automation scripts.

*Manager scripts*

The manager scripts perform two fundamental roles. The first of these is the management of the database and the tool testing process, through tasks such as initiating a bank of virtual machines (see Requirement 10, Table 2). The second role is the movement of software tools and malware into the VMs and extraction of log files created out of these environments.

*The Oracle*

Due to the lack of any theoretical or easily determined 'ground truth', the MATEF determines the expected quantity of artefacts from an independent source (see Requirement 14, Table 2). The random nature of the artefacts generated by malware is such that the reported expected value is little more than an approximation of the 'ground truth'. This source, referred to as the 'Oracle' could be conceivably be any one of a number of online environments, such as that provided by F-Secure (2011) and JoeSandbox (2020).

Unlike online sandbox solutions to analyse malware, use of offline tools enables the investigator greater control of the test environment. Control measures include the configuration of virtual machines and the ability to run tests repeatedly over extended periods of time to identify predictable artefacts. The investigator can also control the distribution of potential personal identifiable information that may be hard coded into a custom-built malware binary. This mitigates the risk whereby malware authors may become alerted to an ongoing investigation by publishing such binaries to a public online platform (C. H. Malin et al., 2008).

*Test environment*

The test environment is managed via automated scripts and enable multiple tests to be run in parallel and thus reduce the time required for large scale tool testing. In addition, this improves the statistical power (and hence the statistical significance) of the results (Smith, 2012) to address the anticipated variability of the malware under analysis.

*Internet simulation*

The provision of network services (see Requirement 11, Table 2) provides the MATEF with an added level of realism to malware running within the Test Environment. Lee et al. (2019) report that as of 2017 over 90% of malware has been found to contain DNS network traffic.

It is important this network provision is simulated to minimise any risk of the malware stealing any data or committing any unauthorised access to other networks (see Requirement 1, Table 2). Requests and responses should be passed to and from common network services that are exposed to the test environment through the component.

Logs of network activity, together with those generated by the tool under test, form a significant product of the test environment and feed into the analysis component.

*Analysis component*

In order to undertake analysis of a software tool, the analysis component needs to establish three things. The first of these is to establish what the tool is to be compared against. As argued above, this should be the expected quantity ('Expected value') of artefacts observed, as opposed their value.

Secondly, the analysis component needs the capability to extract the number of artefacts observed ('Observed value') by the tool under test from a log file bearing a filename that can be determined programmatically. This will allow multiple log files from different VMs and test runs to coexist (see Requirement 15, Table 2).

A third analysis requirement was that the analysis components must establish an assessment of the difference between the Expected and Observed values (see Requirement 7, Table 2). This is a critical value and contributes to establishing the validation of the tool under test. Variation of this value under repeated testing also provides a measure of the repeatability.

*Summary*

Figure 1 shows how the components described above are combined to form the MATEF, together with the information flows between the components. Note boxes in grey are external components that currently sit outside the MATEF. At present the statistical analysis component is performed using an independent statistical analysis tool. It is envisaged that future development of the MATEF will include a statistical component within the MATEF.

The next section will discuss and evaluate the extent to which the design of the framework has met these requirements.

## 5. DISCUSSION

The discussion below reviews the framework requirements, which are reproduced in Table 4, together with a summary of the discussion points below.

The majority of the requirements are met by the design of the framework, for example by handling the malware binaries via scripts on a closed network, access to the malware is restricted and minimises any accidental cross contamination during testing (Requirement 1). In addition, the use of virtual machines provides a platform for rapidly testing and resetting test environments, addressing Requirement 10, while the use of an Internet simulator addresses Requirement 11 to provide a network service enabled environment, conducive to executing malware. Similarly, the framework enables the use of a variety of operating systems, particularly those that are deemed more vulnerable, such as Windows XP (Requirement 12).

The design of the framework also dispenses with the need to have knowledge of the internal operation of a tool under test, reflecting the real-world practice where practitioners use closed source tools. Instead, the framework provides a means to test the expected output of a tool under test (Requirement 13).

However, there are also some requirements (Requirements 2, 5 and 6) that can only be fulfilled, after the work is released to the community for review and consideration of use in practice. As a result, there is scope for further work to evaluate these (see Further Work in the next section).

The provision for reliability (Requirement 3) is comprised of the four elements of reliability, as defined by R v Lundy ([2013] UKPC 28), see Table 4. For simplicity, these will hereafter be referred to as the 'Lundy requirements' and identified individually as 'L1' through to 'L4', respectively.

These four elements are summarised in Table 4 against Requirement 3. Two of these four have been addressed by the design of the framework and are indicated by check marks '☑' in Table 4. First, the capability for testing a technique (see Requirement 3, L1 in Table 4) is addressed by the design of the framework. Secondly, the rate of error associated with a given software tool observing malware artefacts can be determined through repeated measurements and the use of statistical techniques (see Requirement 3, L3 in Table 4). The remaining two elements of the Lundy requirements in Table 4 (L2 and L4) were not addressed and so are marked with a cross '☒' in Requirement 3 of Table 4. These unaddressed requirements mirror Requirement 5 and 6 of the framework, discussed above.

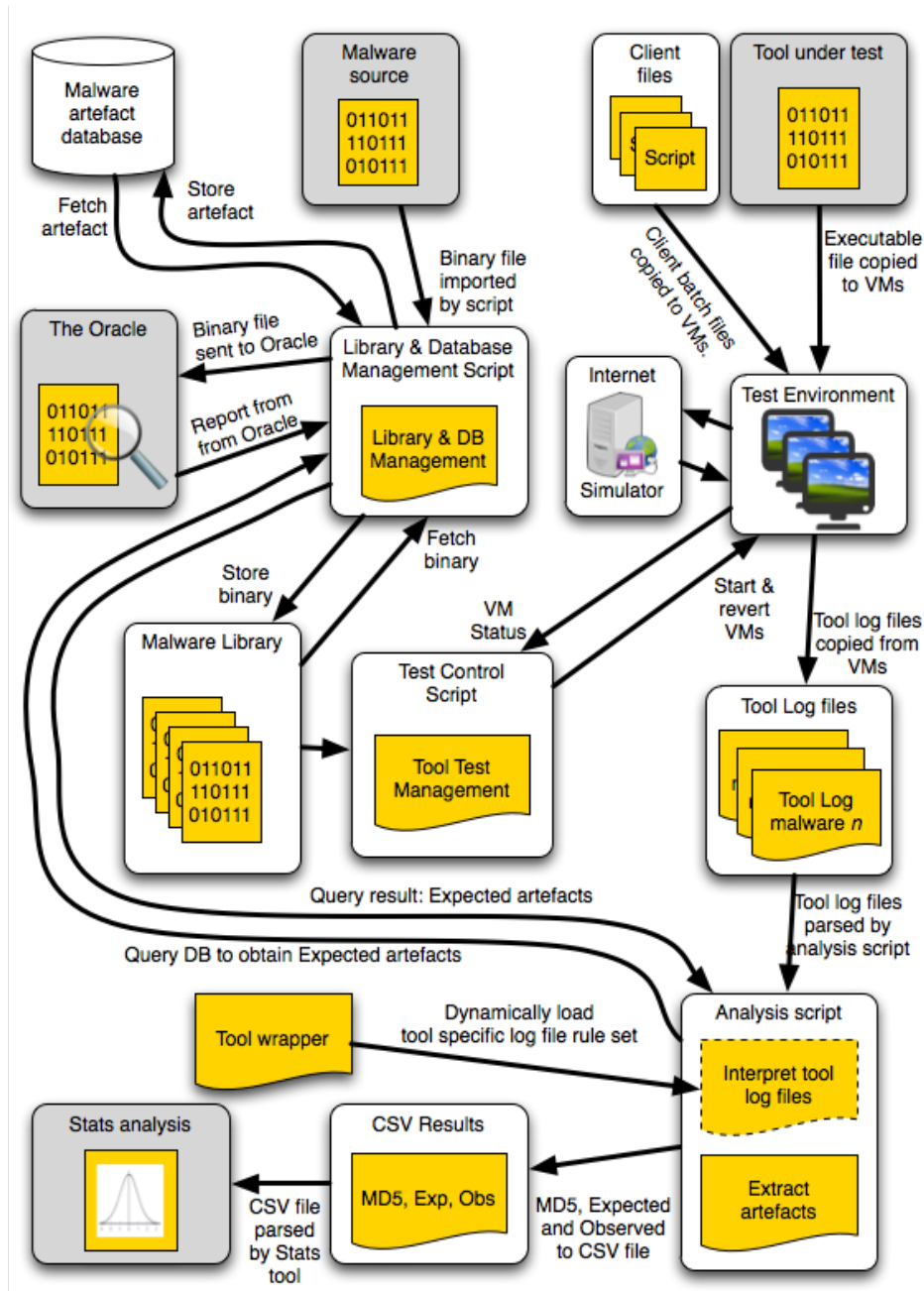

Figure 1 : MATEF components

By publishing the hashes of malware samples sourced from openly shareable resources such as VirusTotal (2010) and using these during the testing of tools, practitioners are able to collaboratively test their tools against the same known and trusted datasets (Requirement 9). Similarly, practitioners can use the same independent source to determine the expected number of artefacts that are generated for a given malware binary (Requirement 14). The design of the framework also allows for these artefacts to be recorded and counted from a disparate range of tools that are subject to testing (Requirement 15).

However, the use of an independent source is not without its problems. Critics of the framework will point to how the requirements for validation (Requirements 4 and 7) have yet to be fully addressed, given the dependence on a third-party tool to provide 'ground truth'. The approach of comparing the results of one tool with that of another (online) one is little more than dual-tool verification. Despite this, the framework's test environment provides a means to test tools under different conditions repeatedly at scale on large numbers of malware binaries. This enables statistical techniques to be applied and thus establish greater confidence in an observed value to a statistically significant degree. This capability facilitates the response to the requirement to provide an estimate for uncertainty (Requirement 8).

Despite the ability to control the frequency and conditions under which the malware is executed locally, it should be noted that the use of online sandboxes to quantify the number of expected artefacts for a given category (i.e., creation, change or deletion) has one notable limitation: rather than providing a representative average quantity, online sandbox tools may only execute samples once and for no more than a maximum time duration before terminating (Bayer, Habibi, Balzarotti, Kirda, & Kruegel, 2009). However, the limitation is attributed to the use of online sandboxes and not of the framework proposed here. The use of online sandboxes was one of convenience, speed and choice (given the variety of online tools available). Additionally, further work is possible here to integrate an offline sandbox, which is under the control of the investigator, to have greater control in how 'ground truth' is established. We discuss this in the next section.

## 6. CONCLUSIONS AND FURTHER WORK

Despite the caveats identified previously, the framework is a versatile platform enabling experiments to be designed according to user-defined protocols. One example is determining the standard deviation from the repeated execution of malware, hence providing an estimate of uncertainty. Indeed, Hubbard (2014, p. 162) points out that where there is a lot of uncertainty in a quantity, then very little data is needed to reduce the uncertainty significantly. Hence, producing an estimate of the expected number of artefacts to be observed significantly reduces the uncertainty in what is expected from subsequent observations.

Support is currently limited to tools that examine file, registry and network artefacts, such as those in the Sysinternals suite (Microsoft, 2020). The tool under test must be capable of being initiated and configured (if needed) via command line. To capture the output of the tool, it must also provide a means to export a log file (in any text-based format). Future work could extend the support to GUI based tools.

The framework is in an early stage of its life-cycle and so further work to critically review and (where appropriate) adopt the framework elsewhere would contribute towards addressing the requirements of the framework relating to critique and general acceptance.

Additional further work could also include the development of an offline oracle. This would provide greater control over parameters, such as execution times and number of runs to better define and hence establish ground truth. A community validated oracle of reference data would provide the greatest level of confidence in the results of the MATEF and could be based on existing projects such as MAEC (Kirillov, Beck, Chase, & Martin, 2010) and ('YARA', n.d.). Improving how ground truth is determined in this way also has the potential to improve the validation requirements of the framework.

There is also scope to harden the virtual environment to minimise detection and evasive behaviour by malware. Furthermore, future work can be done to validate the framework through implementing and testing it with empirical data. Finally, there is room to also engage with practitioners and other stakeholders to gather feedback on the identified requirements and design.

Table 4 : Review of requirements

| # | Requirement | Requirement addressed? |
|---|---|---|
| 1 | Handling of malware and what it may access should be controlled. | Addressed through design:<br>• Scripts on a closed network<br>• Restricted access to the malware<br>• Cross contamination minimised |
| 2 | Output of tested tool must be admissible. | Outstanding. Subject to peer review. |
| 3 | Malware analysis tool output must be reliable | Partially. R v Lundy ([2013] UKPC 28):<br>☑ L1. Technique/theory tested? Yes.<br>☒ L2. Subjected to peer review/publication? No.<br>☑ L3. Reporting the rate of error? Yes<br>☒ L4. Generally accepted? No. |
| 4 | Novel methods must be validated. | Partially. Validation is performed against an authoritative source of ground truth. |
| 5 | The theory/technique should be peer reviewed or published. | Outstanding. Subject to peer review. |
| 6 | Method should be a generally accepted | Outstanding. Subject to peer review. |
| 7 | Tools must be validated. | Partially. Validation is performed against an authoritative source of ground truth. |
| 8 | Estimate of uncertainty | Met through test design: Testing repeatedly at scale to calculate statistical confidence and provide an estimate for uncertainty. |
| 9 | Traceability of reference datasets. | Met through test design: Sharing hashes of malware samples sourced from public malware sources for tool testing enables reproducibility of tests amongst practitioners. |
| 10 | Use a virtual machine. | Met by design: Virtual machines are central to the test environment. |
| 11 | Network service provision. | Met by design: Network services provided via an Internet simulator. |
| 12 | Use vulnerable environment. | Met by design: Virtual machines can use any OS. Windows XP can be used to maximise malware behaviour. |
| 13 | Black box testing approach | Met by design: No internal knowledge of the tool under test is needed. |
| 14 | Metrics:<br>Determine the expected quantity of artefacts from an independent source. | Met by design: The malware artefacts database stores extracted artefacts. |
| 15 | Metrics:<br>Read observed number of artefacts from a variety of tools under test. | Met by design: Observable artefacts captured from log files and translated into a single format for analysis. |